\documentclass[
 reprint,
superscriptaddress,
showpacs,preprintnumbers,
 amsmath,amssymb,
 aps,
 prl,
]{revtex4-1}

\usepackage{graphicx}
\usepackage{dcolumn}
\usepackage{bm}
\usepackage[update,prepend]{epstopdf}

\begin{document}


\title{Spontaneous formation of vector vortex beams in vertical-cavity surface-emitting lasers with feedback }

\author{Jesus Jimenez-Garcia}
\affiliation{SUPA and Department of Physics, University of Strathclyde, Glasgow G4 0NG, Scotland, UK}

\author{Pedro Rodriguez}%

\affiliation{Dpto.\ de Fisica, Universidad de Cordoba, 14071 Cordoba, Spain}
%

\author{T. Guillet}
\affiliation{Laboratoire Charles Coulomb (L2C), Univ. Montpellier, CNRS, Montpellier, France}
\affiliation{SUPA and Department of Physics, University of Strathclyde, Glasgow G4 0NG, Scotland, UK}

\author{T. Ackemann}
\affiliation{SUPA and Department of Physics, University of Strathclyde, Glasgow G4 0NG, Scotland, UK}
\email{thorsten.ackemann@strath.ac.uk}

\date{\today}

\begin{abstract}
The spontaneous emergence of vector vortex beams with non-uniform polarization distribution is reported in a vertical-cavity surface-emitting laser (VCSEL) with frequency-selective feedback. Anti-vortices with a hyperbolic polarization structure and radially polarized vortices are demonstrated. They exist close to and partially coexist with vortices with uniform and non-uniform polarization distributions characterized by four domains of pairwise orthogonal polarization. The spontaneous formation of these nontrivial structures in a  simple, nearly isotropic VCSEL system is remarkable and the vector vortices are argued to have soliton-like properties.
\end{abstract}

\maketitle



Beams with a spatially non-uniform polarization distribution attracted a rapidly growing interest over the last years due to a combination of intrinsic appeal and beauty, novel fundamental aspects in quantum optics and newly enabled applications \cite{zhan09}. They are usually referred to as `vector vortex beams' \cite{milione11,souza14} and possess a circular symmetric intensity structure, most often in form of a doughnut, combined with a spatially non-uniform polarization field. The polarization structure can possess cylindrical symmetry as in beams with radial, azimuthal or spiral polarization  (`cylindrical vector beams' \cite{zhan09,milione11} in a narrow sense). Beams with hyperbolic polarization structure are referred to as `anti-vortices' \cite{maurer07} or `$\pi$-vortices' \cite{milione11}.
They have advantages and applications (see \cite{zhan09} for a review) in engineering tight focusing \cite{dorn03}, micro-machining \cite{allegre12}, optical trapping \cite{kozawa10,roxworthy10}, simultaneous spectroscopy of multiple polarization channels \cite{fatemi11} and beam transformation in nanophotonics \cite{tischler14}. Particularly exciting is the realization that the correlation between spatial and polarization degrees of freedom might open up novel schemes for the use in quantum optics \cite{qian11,kagalwala13,rafsanjani15,ndagano15} and sensing \cite{berg-Johansen15}. Concepts of singular optics have also significant impact outside of optics as recently ideas were transferred to the fields of electron and plasma beams \cite{harris15,bandyopadhyay15}.

Hence, considerable effort was spent on creating these unusual polarization states \cite{zhan09}, relying on a substantial engineering effort based on specialized equipment as tailored laser resonators \cite{fridman08,senatsky12}, meta-surfaces and spatially varying wave plates \cite{machavariani07,liu14,bauer15}, Mach-Zehnder interferometers \cite{roxworthy10}, modal control in few-mode fibers \cite{ramachandran13,ndagano15}, spatial light modulators \cite{maurer07}, tailored Fresnel reflection from glass cones \cite{radwell16},  and  polariton microcavities \cite{lagoudakis09,manni13,sala15}. In contrast, we demonstrate the spontaneous emergence of these structures in a conceptually simple system, a vertical-cavity surface-emitting lasers (VCSEL) with frequency-selective feedback. We will demonstrate the emergence of an anti-vortex as well as spiral and radially polarized vortices, depending on the anisotropies of the feedback scheme. The possibility of a compact and highly adjustable source of variable polarization profile provides potential for subsequent use in applications, in addition to the intriguing fact that such a complex polarization structure is emerging spontaneously. Here the field of singular polarization optics meets soliton science and VCSEL technology.
\begin{figure}
 \resizebox{\hsize}{!}{\includegraphics{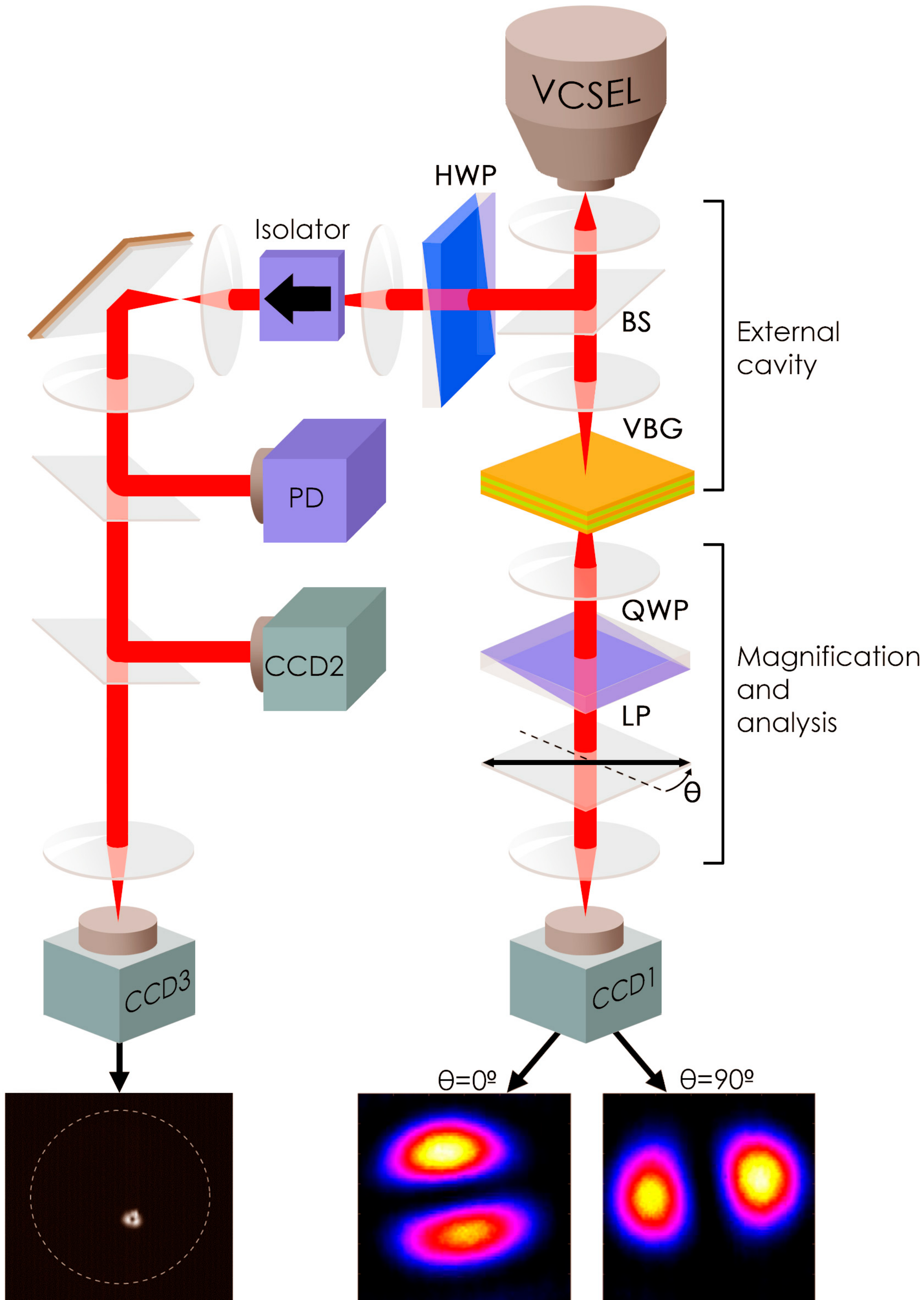}}
  \caption{(Color online) Experimental setup: A volume Bragg grating (VBG) provides frequency-selective feedback to a VCSEL.
  The two lenses forming the self-imaging external cavity have focal lengths of $f_{1}=8$ mm and $f_{2}=50$ mm.
  BS: beam sampler, HWP: half-wave plate; CCD: charge-coupled device camera, PD: photo-detector, LP: linear polarizer, QWP: quarter-wave plate. The right arm is used to measure the spatially resolved Stokes parameters at high magnification (CCD1, example images for the polarization resolved data for the anti-vortex state are shown), the left arm monitors power (PD) and near (CCD3) and far field (CCD2) distributions of potentially the whole laser.}
  \label{setup}
\end{figure}

Vortex solitons were theoretically predicted in lasers for quite a while \cite{fedorov03,rosanov05,paulau10} and observed in VCSELs with saturable absorption \cite{genevet10a} or frequency-selective feedback \cite{jimenez13}. To our knowledge, polarization properties were not considered before in the literature.
Vector vortex solitons in single-pass propagation schemes are predicted to exist for self-defocusing nonlinear media \cite{ciattoni05}, but are known to be unstable for self-focusing media \cite{ishaaya08}, similar to the quasi-scalar case \cite{rosanov94,tikhonenko95,firth97}, although less unstable \cite{bouchard16}.

\begin{figure} 
  \resizebox{\hsize}{!}{\includegraphics{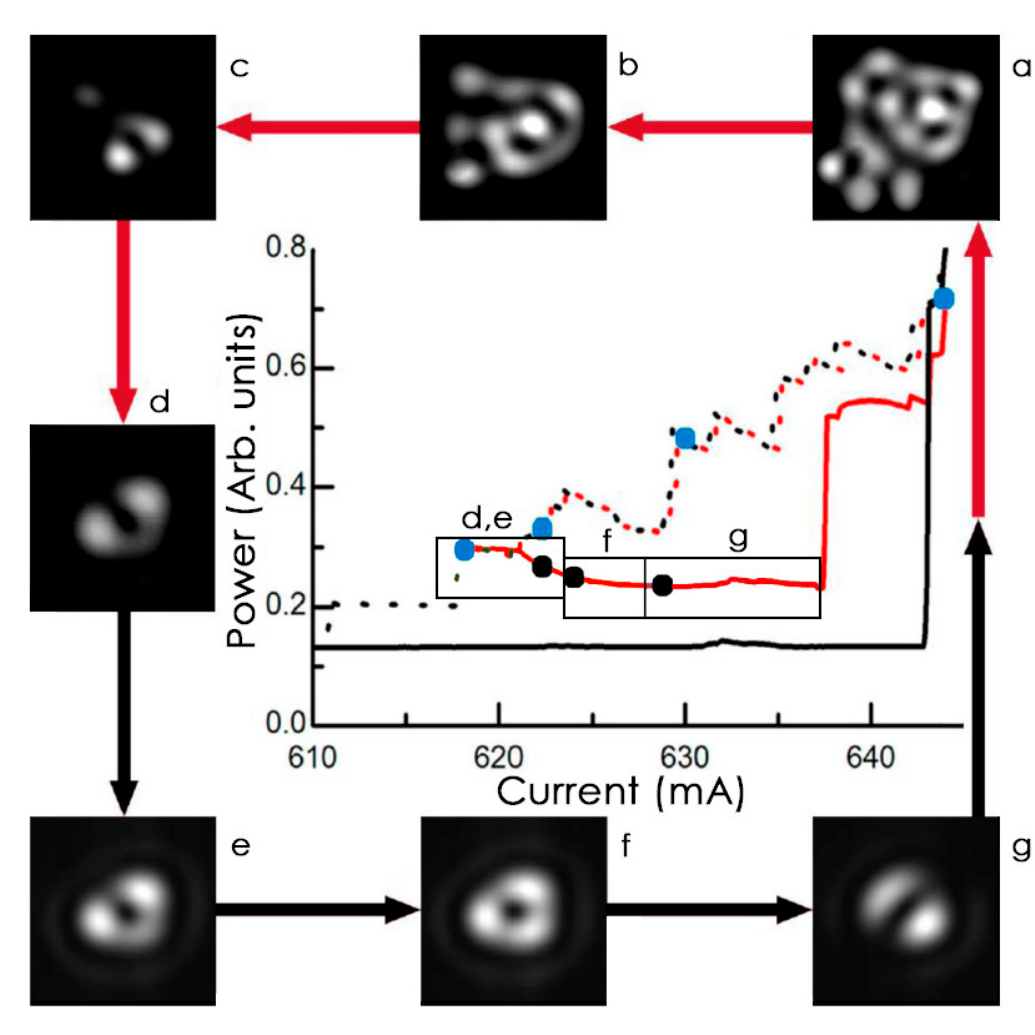}} 
  \caption{(Color online) LI-characteristic obtained by monitoring the output power of the VCSEL via the BS and a linear polarizer oriented at -45$^\circ$. The current is increased from 610~mA in small steps of 0.1~mA till 644~mA (solid black line) and then lowered again till 610~mA (dotted black line) demonstrating bistability between lasing and non-lasing solutions. If the down-scan is stopped at 618~mA and the current increased again till 644~mA, the solid red curve is obtained. The down-scan from the endpoint (dotted red curve) shadows the original downscan (dotted black). e, f, g) Spatial intensity distributions of interest for increasing current are marked by the black dots and displayed in the lower row of images showing, from left to right, the 4-domain  vortex beam (e), the homogeneously polarized vortex beam (f) and the anti-vortex (g) beam. The rectangles indicate the intervals in which the displayed structures exist. The upper row of images (a-d) shows the structures obtained for decreasing current with current values denoted by the blue dots. The structure at the  initialization point at 618~mA is an inhomogeneously polarized 4-domain structure  displayed in (d).}
  \label{fig:IL2}
\end{figure}

By analogy with the quasi-scalar case \cite{fedorov03,rosanov05,paulau10,genevet10a,jimenez13,crasovan00,mihalache08a} one might anticipate also the possibility of stabilization of vector vortices in the flow-equilibrium of driven dissipative systems like cavities. VCSELs are attractive for this kind of studies as they allow a huge variety of spatial \cite{changhasnain91,scheuer99,hegarty99,schulz-ruhtenberg09,yadin14} as well as polarization \cite{changhasnain91,sanmiguel95b,woerdman97,exter98,ackemann01c,schulz-ruhtenberg09,virte12} states due to their high Fresnel number and nominal circular symmetry. Theoretically, vector vortex beams were predicted for VCSEL modes in \cite{prati97}, but never experimentally observed. We are not aware of any experimental or theoretical work on vector vortex solitons, but indications for non-trivial polarization states were found for fundamental solitons \cite{averlant16}. The present letter first focuses on the generation of anti-vortex solitons that present a characteristic hyperbolic polarization pattern. The generation of spiral and radially-polarized vortex solitons will be discussed afterwards.

The experimental setup is very similar to the one utilized in previous works on laser solitons (LS) \cite{radwell09,jimenez13,jimenez16,ackemann16}. The VCSEL used is a large aperture device  with a diameter of 200 micrometers emitting in the 980~nm region \cite{grabherr98,grabherr99,miller01,jimenez17s}. A Peltier element with a feedback circuit is used to stabilize the VCSEL temperature at $20^\circ$C.

The output of the VCSEL is coupled via an afocal telescope onto the frequency-selective element, a volume Bragg grating (VBG).  The VBG has a narrow-band reflection peak of 95\% at $\lambda_{g} = 978.1$ nm, with a reflection bandwidth of $0.1$ nm full-width half-maximum (FWHM). For monitoring the output a wedged glass plate with an uncoated facet at the front and an anti-reflection coated facet at the back serves as an outcoupling beam sampler (BS). The Fresnel reflection provides a polarization anisotropy in the external cavity (discussed to be important later) giving a reflectivity of 10\% for s-polarized light and 1\% for p-polarized light. Note that the polarization asymmetry is much smaller (1:1.1) in transmission. Via a half-wave plate and an optical isolator, polarization resolved light-current (LI) characteristics as well as near and far field intensity distributions can be obtained with a photo-diode and CCD cameras, respectively. The use of the intra-cavity BS allows measurements also without feedback. However, the main results for the polarization distributions are obtained by observing after the VBG as the intra-cavity polarization state can be accessed directly from there. The light that goes through the VBG is re-imaged onto another CCD-camera (CCD1) by a telescopic systems, providing enough magnification to resolve accurately the different polarization zones. Within the collimated range between the two lenses of the  telescope a linear polarizer (LP) and a quarter-wave plate (QWP) are used to measure the spatially resolved Stokes parameter.

\begin{figure}
    \resizebox{\hsize}{!}{\includegraphics{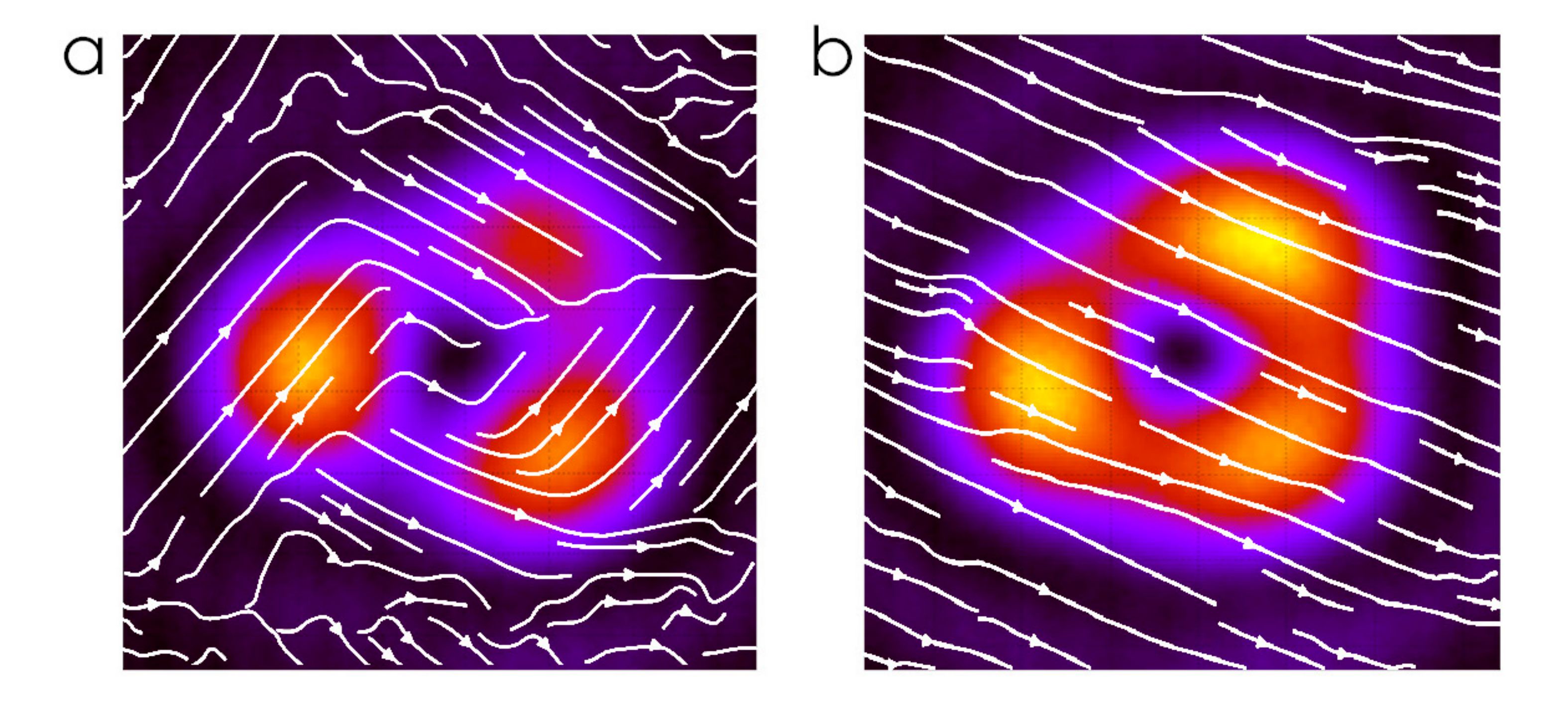}}
  \caption{(Color online) Total intensity S$0$ (in pseudo-colors ranging from black for background via blue and red to yellow for maximum intensity) and polarization streamline diagram (white lines) calculated from the Stokes parameters for the inhomogeneously polarized vortex structure at 622~mA (a) and the homogeneously polarized vortex at 624~mA (b).}
  \label{fig:homogeneo}
\end{figure}

\begin{figure}
\resizebox{\hsize}{!}{\includegraphics{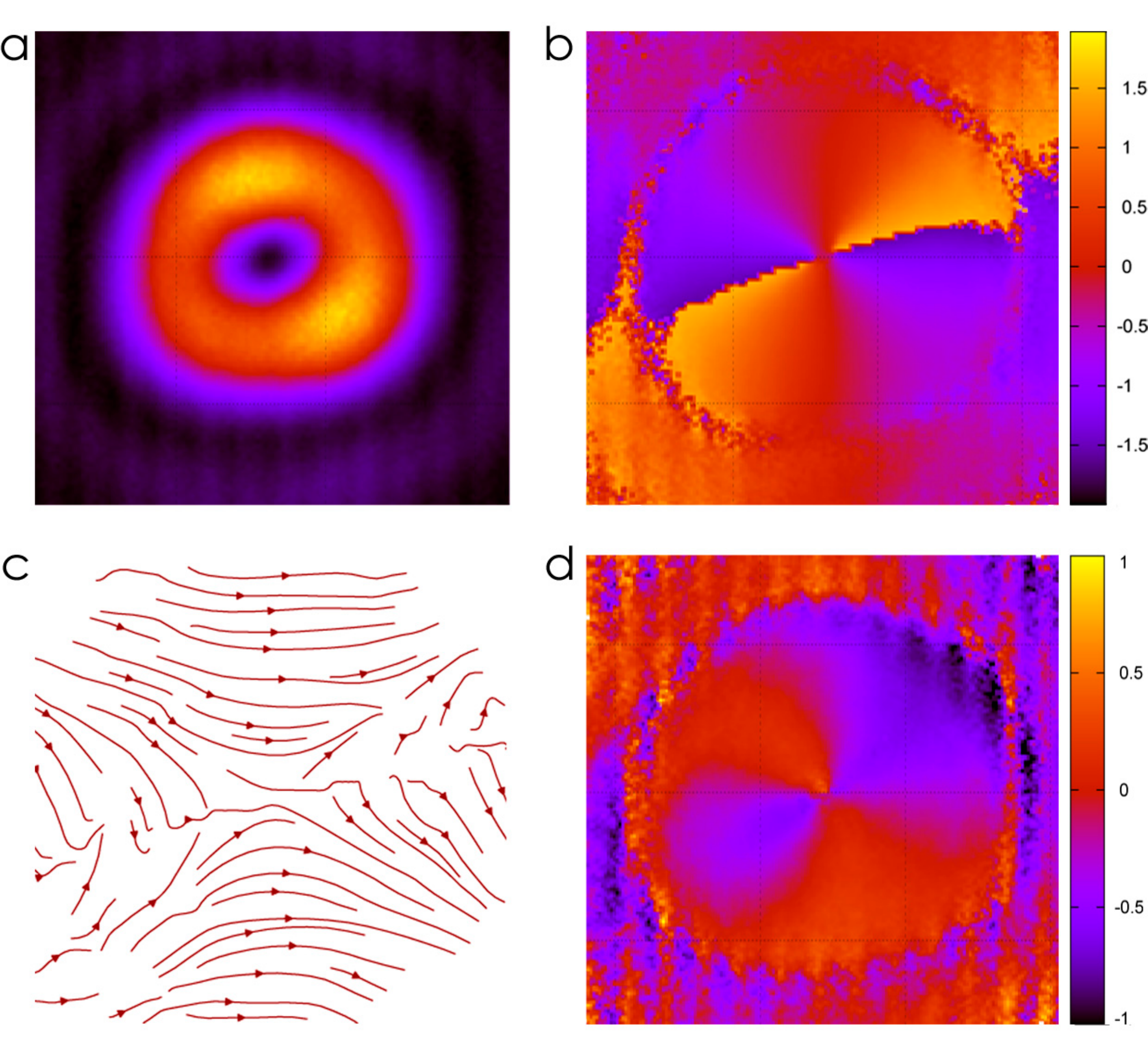}} 
  \caption{(Color online) Polarization configuration for the anti-vortex case (628.3~mA): Total intensity $S0$ (a), local polarization angles (in radians) in pseudocolor coding (b), polarization streamline diagram (c) and the spatial distribution of the circular polarization degree ($S3$ parameter, d).}
  \label{fig:avjfinal}
\end{figure}

When the current supplied to the VCSEL is increased (Fig.~\ref{fig:IL2}), emission stays at spontaneous emission level until it switches up abruptly to a fairly complicated structure (Fig.~\ref{fig:IL2}a) at 643.1~mA (solid black line). Decreasing the current again reveals significant hysteresis (dotted black line). The structures simplify (Fig.~\ref{fig:IL2}b-d) and each transition is abrupt. The structure on the plateau in region d
is a two peak structure with an added arc (Fig.~\ref{fig:IL2}d). The structure between 617.7-610.9 mA is a single bright spot, the fundamental soliton, not shown here. The main effect of increasing the current is actually not an increase in gain but a decrease of the detuning between VCSEL and VBG resonances via Ohmic heating \cite{jimenez16,ackemann16}. At the point of the abrupt switch-on, the detuning between the VCSEL and the VBG is so small that a further nonlinear detuning shift via the carrier (and hence intensity) dependent refractive index \cite{henry82} can bring the two into resonance. Once obtained this state can be sustained, if the current is decreased, respectively the detuning increased, again. This leads to the observed hysteresis or optical bistablity \cite{lugiato84}.

If the down-scan is stopped at 618~mA and the current increased again, the structures of interest for this letter are obtained. Their power is about 100~$\mu$W. By measuring the Stokes parameter after the VBG, the polarization state of each structure is characterized. A typical example
for the states obtained in regions (d,e) is displayed in the left column of Fig.~\ref{fig:homogeneo}. The total intensity ($S0$, Fig.~\ref{fig:homogeneo}a) shows three spots on a triangle with a central dark hole. The polarization is not spatially homogeneous (Fig.~\ref{fig:homogeneo}, see \cite{jimenez17s} for an animation of the polarization state).
We will refer to this inhomogeneously polarized vortex as \textit{`4-domain vortex'}. It is also interesting to note that the polarization distribution of this 4-domain vortex is not random but consists, in the high intensity regions, of four patches which are dominantly either parallel or orthogonal to the  polarization of the homogeneously polarized vortex  (see also the animation in \cite{jimenez17s}) obtained at higher currents. The latter is characterized by a more symmetric structure with three peaks of nearly the same amplitude in the polarization projected image (Fig.~\ref{fig:IL2}f) and the total intensity (Fig.~\ref{fig:homogeneo}b), i.e.\ the polarization state is linear and spatially homogeneous over the beam  (Fig.~\ref{fig:homogeneo}, see also \cite{jimenez17s}). This structure is essentially identical to the vortex soliton observed before at higher temperature \cite{jimenez13}, confirmed to include a phase singularity of order 1 in the dark center. The vortex and fundamental solitons observed at higher temperature have a homogeneous linear polarization and behave quasi-scalar, i.e.\ do not show transitions between different polarization states. We note that the threshold current and thus the available material gain at threshold is much lower   for these quasi-scalar vortices (e.g.\ the threshold is only 200~mA at 46$^\circ$C \cite{jimenez16}) and it is easily conceivable that this makes polarization anisotropies more important. The intensity structure of the observed vortex beams differs from a normal ring vortex solitons by the three-fold azimuthal intensity modulation. Corresponding generalized vortex solitons were predicted \cite{desyatnikov05a} and indications observed \cite{minovich09} in single-pass conservative systems and termed azimuthons. Theoretical predictions exists also for dissipative systems \cite{fedorov03,soto-crespo09} including a simplified model for a VCSEL with frequency-selective feedback \cite{paulau15p}.

If the current is increased further, there is an abrupt transition (see also \cite{jimenez17s}) to a state with two intensity maxima resembling a rotated Hermite-Gaussian H$_{10}$-mode, if viewed through a linear polarizer (Fig.~\ref{fig:IL2}g and insets in Fig.~\ref{setup}). Fig.~\ref{fig:avjfinal} provides a characterization of its polarization state. The total intensity $S0$ (Fig.~\ref{fig:avjfinal}a) resembles a doughnut vortex mode, but the polarization structure is hyperbolic (Fig.~\ref{fig:avjfinal}b, c) with a polarization singularity of order 1. This structures is referred to as \textit{`anti-vortex'} \cite{maurer07} (or $\pi$-vortex beam \cite{milione11}).
As known from the literature \cite{zhan09}, its projection on a specific linear polarization state results in a Hermite H$_{10}$-like structure at a specific angle. A movie in the supplementary material \cite{jimenez17s} and the insets in lower right part of Fig.~\ref{setup} visualize this behavior.  Another important attribute of the anti-vortex structure  is the possibility to transform it to a state with azimuthal or radial polarization by a conventional half-wave plate, as demonstrated in the supplementary material \cite{jimenez17s}.
This opens up, in principle all applications vector vortex beams are considered for.

We mention that a non-negligible amount of circular light component is found in the anti-vortex, making it a full Poincar\'e beam \cite{beckley10}. The results for the $S3$ Stokes parameter for the anti-vortex state is shown in Fig.~\ref{fig:avjfinal}d. The circular component is maximal at the diagonals, i.e.\ where the polarization is radial, and of opposite sign at the two diagonals. Along a ring through the peak intensity, the highest values for $|S3|$ is 0.58 at the diagonal from $-135^\circ$ to $45^\circ$ where also the fractional polarization is lowest (the total fractional polarization varies from 0.8 to 1.04, the deviation of the latter from 1 gives an indication of the accuracy of the measurements). Here $|S3|$ reaches $\sqrt{S1^2+S2^2}$ but is typically much smaller and essentially zero close to the axes. To our knowledge, this is the first example of a spatially non-trivial circular polarization component in a VCSEL, but a similar circular polarization has been measured for an anti-vortex in a polariton condensate \cite{manni13}. A small non-zero $S3$ component was recently found in a fundamental soliton of a VCSEL with injection \cite{averlant16}, but could not be analyzed experimentally with spatial resolution. 	

\begin{figure}
\resizebox{\hsize}{!}{\includegraphics{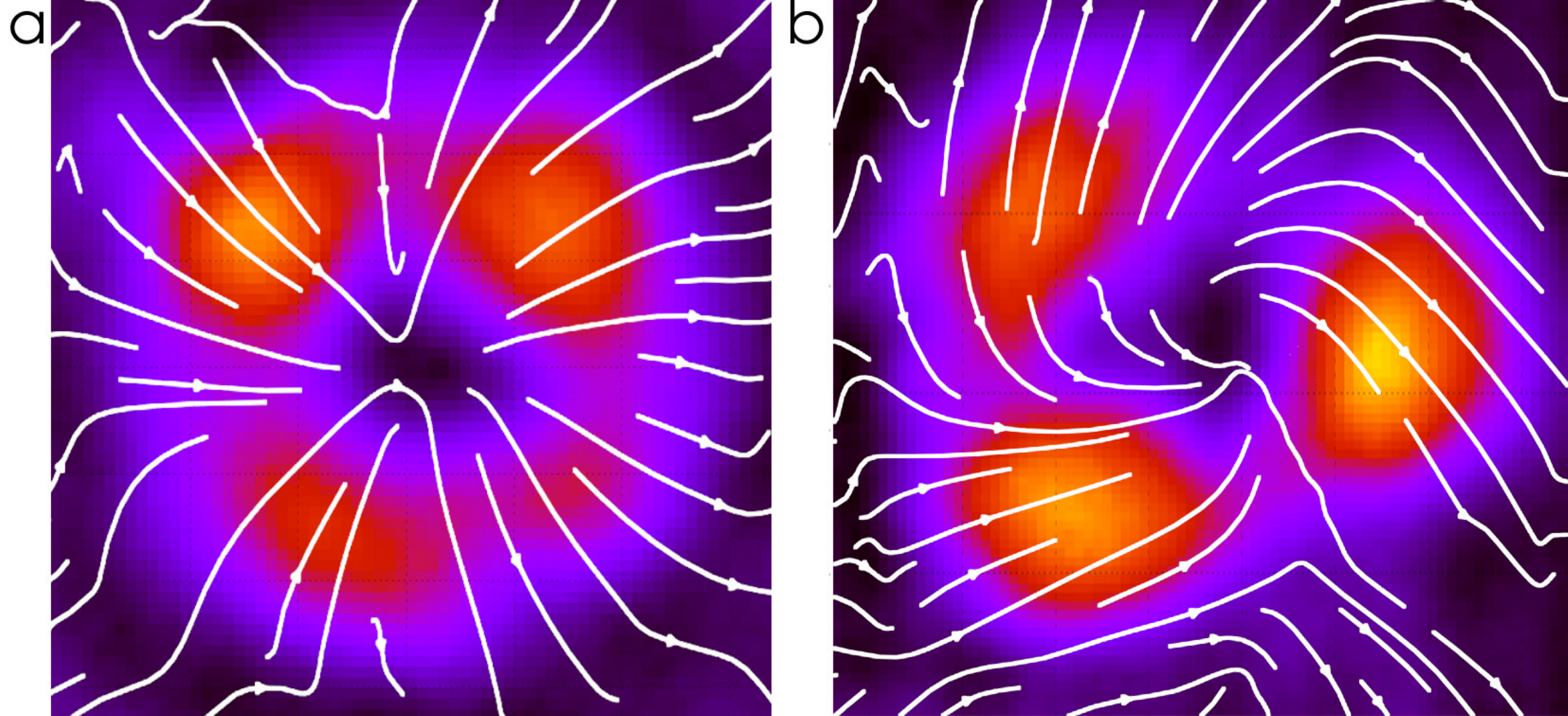}} 
  \caption{(Color online) Polarization configuration for (a) the radially-polarized vector vortex case in a feedback scheme without the BS in the external cavity (625.0~mA) and (b) the spiral vector vortex in a feedback scheme with the BS (626.8~mA): Total intensity $S0$ (pseudo colors) and polarization streamline diagram (white lines).}
  \label{fig:radial_spiral}
\end{figure}

In a further experiment, the beam sampler was taken out of the external cavity making it (nominally) isotropic. In that case, the VCSEL emission exhibits a radially-polarized vortex (Fig.~\ref{fig:radial_spiral}a) with a polarization singularity of order 1 at the core (see also \cite{jimenez17s}), the LI-curve being again very similar to the one of Fig.~\ref{fig:IL2}. Linearly polarized vortices and, in some cases, 4-domain structures are also observed and the transition between them and the radially polarized states is abrupt.
For an interpretation, it is useful to recall that here are two sources of polarization anisotropies in the setup, the intrinsic one in the VCSEL and the one due to the beam sampler in the external cavity. We observe that the more symmetric vector vortex beam (with radial polarization, a cylindrical vortex beam in the classification of \cite{milione11}) is obtained when only the VCSEL anisotropy is involved.  The beam sampler, oriented at about 40$^\circ$ with respect to the VCSEL principal axis, adds a second anisotropy that seems to favor the realization of a state with a lower symmetry, the anti-vortex shown in Fig.~\ref{fig:avjfinal}.


We also mention that a spiral vortex (Fig.~\ref{fig:radial_spiral}b) can be obtained in the system described in Fig.~\ref{setup} (i.e.\ with the BS, see Fig.~S4 in \cite{jimenez17s}). This is not a generic observation, however, and leads to the idea that the different states might coexist and are nearly frequency degenerate but are not linked in parameter space by the standard hysteresis curves.
 Finally, the use of an additional half-wave plate, introduced between the VCSEL and the beam sampler, provides a control of the competition between the anisotropies of the VCSEL and the external cavity, and allows for the robust observation of spiral vortices instead of anti-vortices. This supports the idea that the anisotropies decide which state out of the whole possible and nearly degenerate vector vortex states is stable and accessible.

The theoretical analysis in \cite{prati97} indicates the possibility of the existence of the three types of vector vortex beams that we observe (anti-vortex, spiral vortex and radially polarized vortex) in a free running VCSEL. Perfect cylindrical symmetry is best for a large existence range, but they survive modest anisotropies. No detailed analysis of their stability against linear polarized vortices or between different kind of vector vortex beams is given. In our case the analysis is further complicated by the time delayed frequency-selective feedback and beyond the scope of this letter.

The anti-vortex states share many properties with high-order dissipative solitons as the  localization to a small region in a broadly pumped plano-planar cavity, their coexistence with the non-lasing zero background and the abrupt emergence. Unfortunately, in real VCSEL devices, the detuning fluctuations due to monolayer fluctuations in the growth process are already so large that solitons are typically pinned at certain locations \cite{tanguy08,genevet08,barland02,pedaci08,pedaci08a,ackemann12}. The existence of these traps typically do not alter the properties of the solitons, as shown explicitly for the fundamental solitons in the system under study \cite{jimenez16}. Nevertheless, a clear corroboration of the soliton character comes usually from theoretical confirmation of the stability of the experimentally observed soliton state, as it was the case for the quasi-scalar vortex solitons (azimuthons) obtained at higher ambient temperatures in our system \cite{jimenez13,paulau15p}. As the carrier density is determined by the total power (S0) only and is controlling the nonlinear refractive index, self-guiding is also expected for vector vortex beams.
A detailed theoretical treatment is currently beyond our resources but we hope that this letter triggers theoretical efforts to this effect.

To summarize, playing with the anisotropies via presence or absence of a beam sampler misaligned to the VCSEL principal axis, different  realizations of vector vortex beams can be accessed. The observations suggest that all solutions are likely to exist somewhere in parameter space and are nearly degenerate but their accessibility depend on details of the realization of the anisotropies. A first attempt to analyze the crucial role of anisotropies is present in the work of Prati et al.~\cite{prati97}, but our results call for a more complex theoretical investigation taking into account misoriented principal axes and the external cavity feedback.
The spontaneous emergence of non-trivial polarization structure like the anti-vortex  and the radially polarized vortex is a very remarkable feature, independent of its clear identification as a high-order dissipative vector soliton. In many respects, the setup described here is much simpler than typical schemes to create vector vortex beams and thus might open up new opportunities for applications, but obviously a lot of technological development is needed to turn this scheme into a robust working system.

\section*{Acknowledgements} We are grateful to Gian-Luca Oppo and William J.\ Firth for useful discussions and to Roland J\"ager (Ulm Photonics) for supplying the devices. J.J.\ gratefully acknowledges support from CONACYT.  P.R. acknowledges to Universidad de Cordoba for his support and Jacobo Muniz-Lopez for his help in the configuration of some figures. We are grateful to the referees for very useful comments.


%

\end{document}